# A substructure inside spiral arms, and a mirror image across the Galactic Meridian

Jacques P. Vallée

National Research Council Canada, Herzberg Astronomy & Astrophysics,
postal address: 5071 West Saanich Road, Victoria, B.C., Canada V9E 2E7
emails: jacques.p.vallee@gmail.com jacques.vallee@nrc-cnrc.gc.ca

## Abstract

While the galactic density wave theory is over 50 years old and well known in science, whether it fits our own Milky Way disk has been difficult to say. Here we show a substructure inside the spiral arms. This substructure is reversing with respect to the Galactic Meridian (longitude zero), and crosscuts of the arms at negative longitudes appear as mirror images of crosscuts of the arms at positive longitudes. Four lanes are delineated: mid-arm (extended $^{12}CO$ gas at mid arm, HI atoms), in-between offset by about 100 pc (synchrotron, radio recombination lines), in between offset by about 200 pc (masers, colder dust), and inner edge (hotter dust seen in Mid-IR and Near-IR).

## 1. Introduction

Elsewhere, in nearby external disk galaxies, the search for offsets between chemical tracers of spiral arms has been positive in some spiral galaxies, yet negative in some other galaxies, while it is ambiguous in other spiral galaxies such as in the giant galaxy M51with negative and positive results (Louie et al 2013).

What is it in the Milky Way galaxy? The search for a substructure inside spiral arms in our Milky Way galaxy has been difficult, as we cannot easily observe in the optical domain far inside the galactic disk nor can we go above the galactic disk. Here, we use the tangent to each spiral arm, found when scanning the galactic disk in galactic longitudes, using different chemical tracers in the radio and far infrared domains. When looking through the tangent of a spiral arm, a peak intensity in that tracer appears. This observing technique cannot be done by us in any other galaxy, although a similar study could in principle be done in edge-on spiral galaxies.

Most of the chemical tracers employed to get the tangents for spiral arms inside the Milky Way are detected at radio wavelengths, although not all. Published arm tangents are listed in Vallée (2014a – his statistical table 3 has 43 arm tangents from 10 different tracers) and in Vallée (2014b – his statistical table 1 has 63 arm tangents from 14 different tracers), while in Appendix A here the statistical table 3 has 88 arm tangents from 18 different tracers, using 125 individual observations listed in Tables 4 to 10. These papers discovered for the first time that the dust lane in the Milky Way's spiral arms was offset from the extended gas in the middle of the arm, notably by a few degrees in longitude from the center of the arm where the extended CO gas is mostly located. This CO tracer is the low-excitation J=1-0, low temperature (near 10 K), low-angular resolution (near 9'), narrow line (near 2 km/s) at 115 GHz, integrated over a velocity range commensurate with a spiral arm, from the Columbia survey.

Very recently, we did some analysis on the average offset between the arm edge and the arm center in the Milky Way galaxy. The average offset from the hot dust lane to the CO gas in mid-arm was found near 340 pc ± 56 pc, when averaging over all spiral arms with an observed tangent from the Sun (Fig. 2 in Vallée 2014b). This is now well established (a 6 sigma result).

Also, the average offset from the methanol maser to the CO gas in mid-arm was found *separately* for the group of so-called 'major' arms (Crux, Scutum. Perseus – 444 pc) and for the group of so-called 'minor' arms (Sagittarius, Carina, Norma – 401 pc), giving a width radio = 1.1 ± 0.2, essentially unity within the errors (Fig.3 in Vallée 2014a), and thus contrary to the putative existence of alternating 'major' (wide) and 'minor' (narrow) arms in the Milky Way.

Any arm substructure could be useful when attempting to place galactic filaments (very long, very thin, infrared dark clouds) in spiral arms, either at the spiral arm middle (Zucker et al 2015 – their fig.2), or else at the inner edge of spiral arm (Wang et al 2015 – their Sect. 4 .2), or outside in the inter-arm region (Ragan et al 2014 – their fig 4).

## 2. Updated Analysis – across the Galactic Meridian

Here, we analyse the offset of the dust to the CO gas peak, but *separately* at negative longitudes and at positive longitudes from the Galactic Center (longitude $0^o$).

**Table 1** shows that the arm tangents for hot dust are offset inward from the arm tangents for the $^{12}$CO, thus closer to the direction of the Galactic Center. Here the average offset between the red lane (dust) and the blue lane (mid arm, $^{12}$CO) is about 315 pc, inside the solar orbit around the Galactic Center.

**Figure 1** illustrates graphically this point. Galactic quadrants I corresponds to positive longitude l from 0 to 90 degrees, quadrant II to longitude l from 90 to 180, quadrant III from 180 to 270 degrees, and quadrant IV from 270 to 360 degrees (equivalently, from -90 to 0 degrees).

Although in Table 1 and Figure 1 some arms in quadrant IV appear to have a counterpart in quadrant I, these arms are not physically linked; the $^{12}$CO gas in Crux-Centaurus arm at -51$^{o}$ is not linked to the $^{12}$CO gas in the Sagittarius arm at +51$^{o}$; the hot dust in the Norma arm at -28$^{o}$ is not linked to the hot dust in the Scutum arm at +29$^{o}$.

A map of the Milky Way disk shows that the Crux-Centaurus arm is continuously linked to the Scutum arm, while the Carina arm is continuously linked to the Sagittarius arm, owing to a spiral pitch angle (a deviation from a circle) of about -13$^{o}$ (Vallée, 2015 – his fig.1).

## 3. New analysis – a mirror image across the Galactic Meridian

Here, we look for the presence of further lanes, in between the red (dust at arm's inner edge) lane and the blue ($^{12}$CO 1-0 gas at arm's middle), using an enlarged catalog of arm tangents (Appendix A, with 88 arm tangents from 18 different arm tracers).

The 4-arm structure in the Milky Way may encompass some substructures inside each spiral arm. We define here a 'substructure' when each different tracer is seen the same general arm at its own distance between the dust lane and the $^{12}$CO 1-0 lane, but reversing on the other side of the Galactic Meridian (mirror image).

**Table 2** shows the offset between two tangents, that of the $^{12}$CO tracer direction as seen from the Sun and that of a specific arm tracer (molecule, atom, electron, excited gas, hot and cold dust). The angular offset is converted to a linear separation through its known distance from the Sun.

**Figure 2** shows the mean arm crosscut, showing the offset for each tracer from the mid arm. They are averaged separately for the arms left of the Galactic

Meridian (quadrant IV) in Fig.2a, and right of the Galactic Meridian (quadrant I) in Fig.2b.

A blue lane encompasses the mid arm (including $^{12}CO$). A red lane encompasses the arm's inner edge (including the hot dust). In between, we observe a 'substructure', namely one sees a green lane (near 100 pc, including relativistic synchrotron electrons, and radio recombination lines), and an orange lane (near 200 pc, including FIR [CII] and [NII] lines, radio masers, and colder dust at Extreme InfraRed).

At first glance, these two quadrants are mirror images – each lane (blue, green, orange, or red) is located at about the same place in linear separation. The mean arm width is about the same, with the red dust at 378 pc (left) and 340 pc (right), giving a ratio of 1.1 ±0.3, thus nearly equal within the errors.

## 4. Arm width with galactocentric distance

Finally, we reassess the possible enlargement with galactocentric distance of the arm half-width (offset between the red lane and the blue lane).

From the Notes in Table 1, using offsets from hot dust to $^{12}CO$ 1-0, one has a mean arm half-width for the Sagittarius and Scutum arms of 280 ±69 pc in galactic quadrant I at a mean galactic distance of 4.5 kpc, which is smaller than the mean arm half-width for the continuation of these two arms into the other galactic quadrant IV (Carina and Crux-Centaurus) of 305 ±44 pc at a mean galactic radius of 5.5 kpc, giving a ratio of 1.1 ±0.3, for a mean galactocentric distance ratio of 1.25 (see bottom of Table 1).

From Table 2, using offsets from cold dust (EIR - 870μm) versus $^{12}CO$ 1-0, one has a mean arm half-width for the Sagittarius and Scutum arms of 136 ±37 pc in galactic quadrant I at a mean galactic distance of 4.5 kpc, which is smaller than the mean arm half-width for the continuation of these two arms into the other galactic quadrant IV (Carina and Crux-Centaurus) of 225±27 pc at a mean galactic radius of 5.5 kpc, giving a ratio of 1.6 ±0.5, for a mean galactocentric distance ratio of 1.25.

A statistical mean of these two observed arm ratios gives a rough value of 1.35 at a distance ratio of 1.25.

Table 1 – Arm tangents to two arm tracers (hot dust; extended $^{12}CO$ gas), in galactic longitude, using a Sun to Galactic Center distance of 8.0 kiloparsecs

| | $^{12}CO$ tangent at gal. longit. (degrees) | dust tangent at gal. longit. (degrees) | angular difference (degrees) | linear difference (parsecs) | Note - |
|---|---|---|---|---|---|
| Carina arm | -79 | -75 | 4 | 350 (at 5kpc) | 1 |
| Crux-Cent. Arm | -51 | -48.5 | 2.5 | 260 (at 6kpc) | 1 |
| Norma arm | -31 | -28 | 3 | 370 (at 7kpc) | |
| Perseus-start arm | -23.5 | -21 | 2.5 | 350 (at 8kpc) | |
| Scutum arm | +33 | +29 | 4 | 350 (at 5kpc) | 2 |
| Sagittarius arm | +51 | +48 | 3 | 210 (at 4 kpc) | 2 |
| Mean and s.d.m. | - | - | 3.2 ± 0.3 | 315 ± 26 | |

Note 1: the mean offset for Carina and Crux-Centaurus is 305 ± 44 pc at a mean solar distance of 5.5 kpc (mean galactocentric distance of 7.5 kpc).
Note 2: the mean offset for Scutum and Sagittarius is 280 ± 69 pc at a mean solar distance of 4.5 kpc (mean galactocentric distance of 6 kpc).

Table 2 – Separation (offset) from the mid arm ($^{12}$CO)

| 1 | 2 | 3 | 4 | 5 | 6 | 7 | 8 | 9 | 10 | 11 |
|---|---|---|---|---|---|---|---|---|---|---|
| | $^{12}$CO mol. pc | ther. el. pc | HI atom pc | relat. el. pc | reco. el. pc | EIR dust pc | FIR [CII] pc | radio maser pc | MIR dust pc | NIR dust pc |
| Carina arm | 0 | 148 | 22 | - | 262 | 253 | 497 | - | 322 | - |
| Crux-Cent. Arm | 0 | -52 | 42 | 52 | 178 | 198 | -52 | - | 157 | - |
| Norma arm | 0 | -49 | -37 | -49 | 110 | 147 | - | 244 | 73 | 449 |
| Perseus-start arm | 0 | - | 0 | 307 | 14 | 140 | 167 | 140 | 446 | 307 |
| Mean | 0 | 16 | 07 | 103 | 141 | 184 | 204 | 192 | 250 | 378 |
| S.d.m . | - | ±66 | ±17 | ±105 | ±52 | ±26 | ±159 | ±51 | ±83 | ±70 |
| Scutum arm | 0 | 78 | 166 | 78 | 183 | 174 | 253 | 384 | 602 | 340 |
| Sagittarius arm | 0 | 105 | -7 | 175 | 91 | 98 | 35 | 35 | - | - |
| Mean | 0 | 91 | 80 | 126 | 137 | 136 | 144 | 210 | 602 | 340 |
| S.d.m . | - | ±31 | ±86 | ±48 | ±46 | ±27 | ±108 | ±173 | - | ±70 |

Note. Column 1 is the spiral arm name, Columns 2 to 11 have the offset, from the $^{12}$CO tangent, of a specific arm tracer tangent, namely the thermal electron, the HI atom, the relativistic synchrotron electron, HII recombining electrons (radio recombination line at 1.4 GHz), the Extreme InfraRed dust (870 microns), the Far InfraRed [CI] and [NII] lines, the radio masers (mostly methanol), the Mid InfraRed dust (60 microns), and the Near InfraRed dust (2.4 microns). The S.d.m. is the standard deviation of the mean.

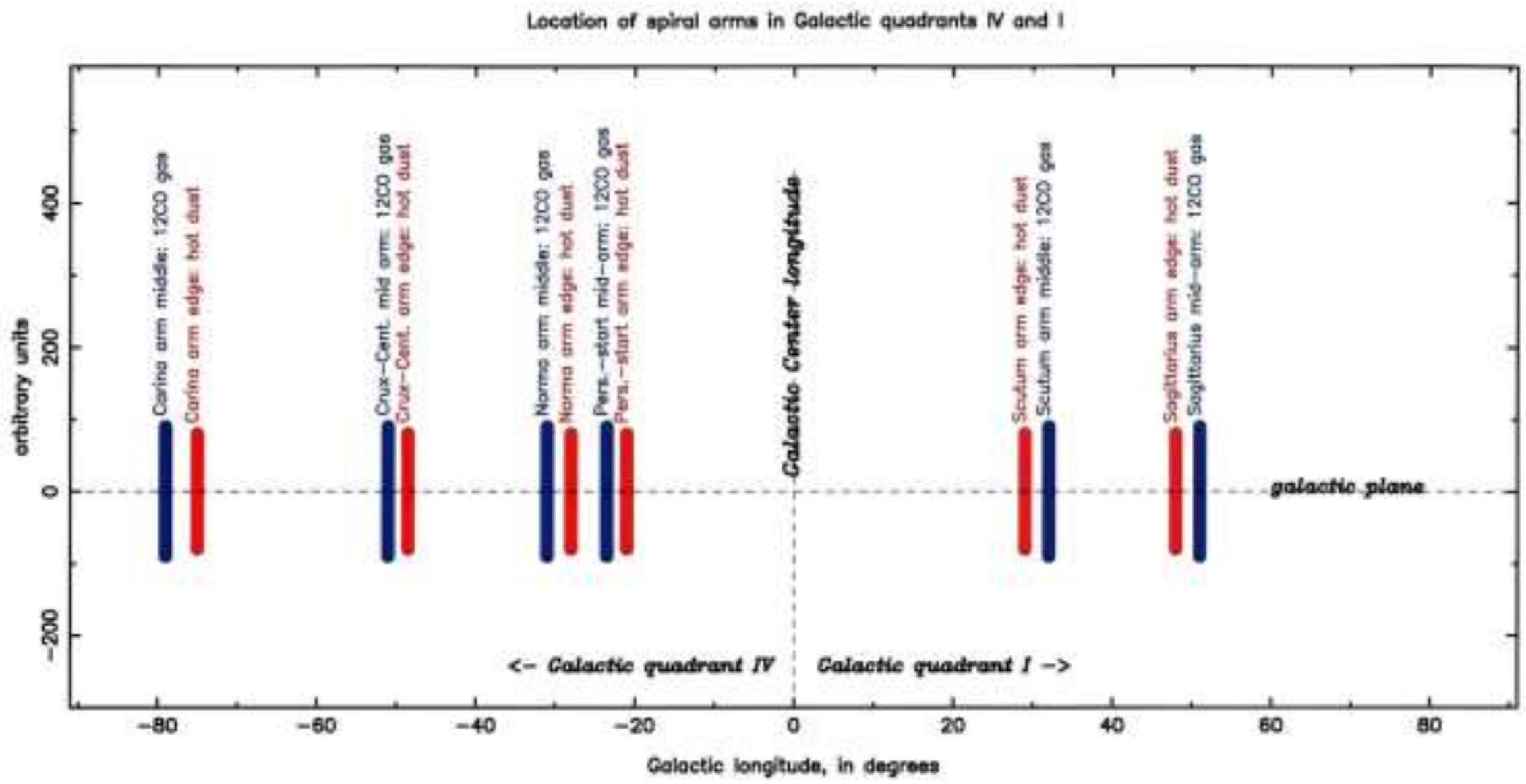

Figure 1. Location in galactic longitude (x-axis) of several spiral arms, indicating the tangent to each spiral arm as seen from the Sun, using two arm tracers ( $^{12}CO$ J=1-0 extended gas, hot dust). An evident reversal of these two tracers is observed as one goes across the Galactic Meridian (located at galactic longitude zero). Vertical axis is arbitrary. In each arm, it can be seen that the red lane (hot dust) is always closer to the Galactic Meridian (located at longitude l =0$^{o}$). Thus the dust lane (red) is to the right of the blue lane (mid arm, $^{12}CO$) when at a negative galactic longitude (quadrant IV), and is reversing to be at the left of the mid arm when at positive galactic longitude (quadrant I).

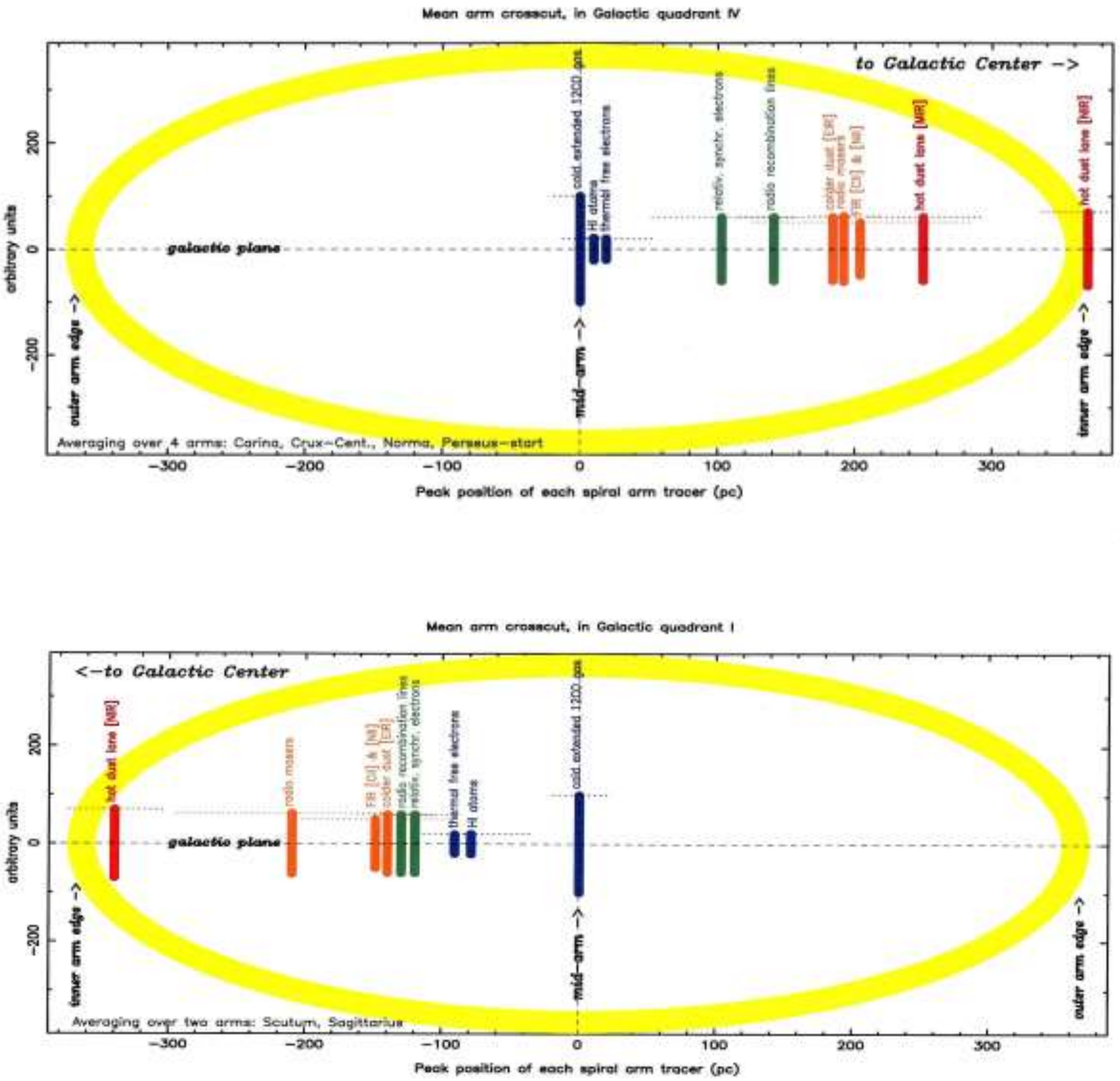

Figure 2. Location of the tangents for several arm tracers, each tracer being averaged over several spiral arms in quadrant IV (Fig.2a) and in Quadrant I (fig.2b). Galactic Center direction (arrow at top) is indicated in each quadrant. These figures are almost mirror images. Blocks of tracers, having similar distances, are shown: hot dust at the inner arm edge (red), mid arm extended $^{12}CO$ J=1-0 gas (blue), and in between blocks near 100pc (green) and near 200pc (orange) as measured from the mid arm. Tangential arm offsets are translated in linear offsets, using the known tangential arm distance (from Table 1). The overall yellow ellipse indicates a very rough limit for the extent of the arm stars.

## 5. Interpretation and Comparison

Not many theories of spiral arms predict a reversal of dust position across the Galactic Meridian, nor of additional substructures inside a spiral arm. A recent review of different theories to produce spiral arms is given by Dobbs and Baba (2014). Here we focus our interpretation using predictions from the density wave theory.

a- Reversal of the inner arm edge with the Galactic Meridian (Section 2):

Density wave theory (Lin & Shu 1964; Lin et al 1969) predicts that the interarm gas, in a circular orbit around the Galactic Center, will overtake a slower moving spiral pattern, and be shocked at the entrance (leading to a hot dust lane), thus the hot dust lane would face inward toward the Galactic Center (below a galactic radius called co-rotation, at about 9 kpc in the Milky Way disk). In the density wave theory, the dust lane corresponds to the predicted location of a shock lane.

In our Milky Way disk galaxy, radio and infrared observations of the arm tangents showed objectively that a reversal between hot dust and mid arm do appear (Figure 1, Table 1), and that a 4-arm structure is needed.

b- Substructures within an arm (Section 3):

Past the shock lane, density wave theory predicts a decrease of gas density and gas temperature, as the orbiting gas penetrates deeper toward the colder arm middle, leading to the prediction of some substructures (regions between the inner arm edge's dust lane and the mid-arm lane in the center). In the density wave theory, the middle of the arm corresponds to the lane with a potential maximum.

In the density wave model of Roberts (1975 – his Fig. 2), along a typical gas streamline around the Galactic Center, a shock (gas density of 5 units) precedes the potential minimum (gas density near 2.5 units), and the orbit brings it to the zero potential (gas density near 1.25 units) and onward to the potential maximum (mid arm; gas density near 0.7 unit), with the gas density being relatively stable afterwards.

The substructure in an arm, predicted by the density wave as a function of the orbit, is theoretical and usually made for a 2-arm spiral model. They should not be expected to be the same, but they should still be close, to the 4 observed substructures in a typical Milky Way's arm.

In the Milky Way disk galaxy, the observational existence of similar gas compositions (substructures) at different offsets from the dust lanes, appears here as some kind of mirror images across the Galactic Meridian (Figure 2, Table 2).

c- Enlargement of an arm width with distance (Section 4):

Density wave theory predicts a weak opening of a spiral arm as a function of galactic distance. In the density wave theory, a small opening of each spiral arm is predicted as a function of increasing radial distance from the Galactic Center.

A 2-arm theoretical map of the Carina-Sagittarius arm by Lin et al (1969 – their fig. 4), drawn for a Sun to Galactic Center distance of 10 kpc and an arm pitch angle near $-6^o$, shows an arm width of 1.9 kpc at the arm tangent to Carina at a galactocentric radius of 8.3 kpc, and an arm width of 1.1 kpc at the arm tangent to Sagittarius at a galactocentric radius of 7.3 kpc, thus giving an arm ratio of 1.7 at a galactic radius ratio of 1.14.

The same theoretical map of the Crux-Centaurus-Scutum arm by Lin et al (1969 – their fig. 4) shows an arm width of 1.2 kpc at the arm tangent to Crux-Centaurus at a galactocentric radius of 5.4 kpc, and an arm width of 0.85 kpc at the arm tangent to Scutum at a galactocentric radius of 4.8kpc, thus giving an arm ratio of 1.4 at a galactic radius ratio of 1.12.

A statistical mean of these two theoretical arm ratios gives a rough value of 1.55 at a distance ratio of 1.13.

## 6. Conclusion

Based on observed arm tangents, we presented statistical results and graphic figures about substructures in the spiral arms in the Milky Way. Substructures (parallel lanes) are observed between the dust lane and the middle of the arm (extended $^{12}CO$ 1-0 gas lane) – see Figure 1 and Table 1.

Furthermore, when comparing a graph of an average of the offsets within spiral arms at negative galactic longitudes (Fig. 2a) with a graph of the offsets within spiral arms at positive galactic longitudes (Figure 2b), we obtain a rough mirror image.

These data tend to support several predictions from the density wave theory, as applied to our Milky Way galaxy. With increasing galactic radius, we statistically deduce a weak enlargement of the offset between dust lane and mid arm.

## Appendix A

Appendix A is an updated compilation of arm tangents, as statistically averaged for each tracer (Table 3), and individually listed for each tracer (Tables 4 to 10). Earlier such catalogues were given in Vallée (2014a; 2014b).

## Acknowledgements.

The figure production made use of the PGPLOT software at the NRC Canada in Victoria, BC. I thank a referee for clarification.

**Table 3 - Catalog of statistical means of arm tangent longitudes for each tracer, in each spiral arm[(a)]**

| Arm Name | Chemical tracer | Tangent [(b)] galactic longitude l | Ang. dist.[(c)] to $^{12}CO$ from $^{12}CO$ | Linear separation inside arm[(cd] | References |
|---|---|---|---|---|---|
| Carina | $^{12}CO$ at 8' | 281.3° | 0° | 0 pc, at 5 kpc | mean in Table 5 |
| | HI atom | 282.1° | 0.8°` | 22 pc | mean in Table 8 |
| | Thermal electron | 283° | 1.7° | 148 pc | Taylor & Cordes (1993 – fig.4) |
| | HII complex | 283.8° | 2.5° | 218 pc | mean in Table 6 |
| | Dust 240µm | 284° | 2.7° | 235 pc | Drimmel (2000 –fig.1) |
| | Dust 870 µm | 284.2° | 2.9° | 253 pc | mean in Table 9 |
| | 1.4GHz RRL | 284.3° | 3° | 262 pc | Hou & Han (2015 – table 1) |
| | Dust 60µm | 285° | 3.7° | 322 pc | Bloemen et al (1990 – fig.5) |
| | FIR [CII] & [NII] | 287° | 5.7° | 497 pc | Steiman-Cameron et al (2010 – sect. 2.1) |
| Crux-Cen-taurus | Old stars (1-8 µm) | 307.3° | - 2.2° | -230 pc | mean in Table 10 |
| | Thermal electron | 309° | - 0.5° | -52 pc | Taylor & Cordes (1993 – fig.4) |
| | FIR [CII] & [NII] | 309° | - 0.5° | -52 pc | Steiman-Cameron et al (2010 – sect. 2.1) |
| | $NH_3$ 1-1 2' cores | 309.1° | - 0.4° | -42 pc | Hou & Han (2015 – table 1) |
| | $^{12}CO$ at 8' | 309.5° | 0° | 0 pc, at 6 kpc | mean in Table 5 |
| | HI atom | 309.9° | 0.4° | 42 pc | mean in Table 8 |
| | HII complex | 309.9° | 0.4° | 42 pc | mean in Table 6 |
| | $^{26}Al$ | 310° | 0.5° | 52 pc | Chen et al (1996- fig.1) |
| | Sync. 408 MHz | 310° | 0.5° | 52 pc | Beuermann et al (1985 – fig.1) |
| | Dust 240µm | 311° | 1.5° | 157 pc | Drimmell (2000 – fig. 1) |
| | Dust 60µm | 311° | 1.5° | 157 pc | Bloemen et al (1990 – fig.5) |
| | 1.4GHz RRL | 311.2° | 1.7° | 178 pc | Hou & Han (2015 – table 1) |
| | Dust 870µm | 311.4° | 1.9° | 198 pc | mean in Table 9 |
| Norma | $^{26}Al$ | 325° | - 3.4° | -415 pc | Chen et al (1996 – fig.1) |
| | HII complex | 326.4° | - 2.0° | -244 pc | mean in Table 6 |
| | $NH_3$ 1-1 2' cores | 327.8° | - 0.6° | -73 pc | Hou & Han (2015 – table 1) |
| | Thermal electron | 328° | - 0.4° | -49 pc | Taylor & Cordes (1993 – fig.4) |
| | Sync. 408 MHz | 328° | - 0.4° | -49 pc | Beuermann et al (1985 – fig.1) |
| | [CII] | 328° | - 0.4° | -49 pc | Velusamy et al (2015 – fig.7b) |
| | HI atom | 328.1° | - 0.3° | -37 pc | mean in Table 8 |
| | $^{12}CO$ at 8' | 328.4° | 0° | 0 pc, at 7 kpc | mean in Table 5 |
| | Dust 60µm | 329° | 0.6° | 73 pc | Bloemen et al (1990 – fig.5) |
| | 1.4GHz RRL | 329.3° | 0.9° | 110 pc | Hou & Han (2015 – table 1) |
| | Dust 870µm | 329.6° | 1.2° | 147 pc | mean in Table 9 |
| | Masers | 330.4° | 2.0° | 244 pc | mean in Table 7 |
| | Dust 240µm | 332° | 3.6° | 440 pc | Drimmell (2000 – fig. 1) |
| | Dust 2.4µm | 332° | 3.6° | 449 pc | Hayakawa et al (1981 – fig.2a) |
| Start of Per-seus | $^{12}CO$ at 8' | 336.8° | 0° | 0 pc, at 8 kpc | mean in Table 5 |
| | HI atom | 336.8° | 0° | 0 pc | mean in Table 8 |
| | 1.4GHz RRL | 336.9° | 0.1° | 14 pc | Hou & Han (2015 – table 1) |
| | [CII] | 337° | 0.2° | 28 pc | Velusamy et al (2015 – fig.8b) |
| | HII complex | 337.2° | 0.4° | 56 pc | mean in Table 6 |
| | Dust 870µm | 337.8° | 1.0° | 140 pc | mean n Table 9 |
| | Masers | 337.8° | 1.0° | 140 pc | mean in Table 7 |
| | FIR [CII] & [NII] | 338° | 1.2° | 167 pc | Steiman-Cameron et al (2010 – sect. 2.1) |
| | Old stars (1-8 µm) | 338.3° | 1.5° | 209 pc | mean in Table 10 |
| | $NH_3$ 1-1 2' cores | 338.4° | 1.6° | 223 pc | Hou & Han (2015 – table 1) |
| | Sync. 408 MHz | 339° | 2.2° | 307 pc | Beuermann et al (1985 – fig.1) |
| | Dust 2.4μm | 339° | 2.2° | 307 pc | Hayakawa et al (1981 – fig.2a) |
| | Dust 60μm | 340° | 3.2° | 446 pc | Bloemen et al (1990 – fig.5) |
| Start of Sagit-tarius | $^{12}$CO at 8' | 344° | 0° | 0 pc, at 7.5 kpc | mean in Table 5 |
| | $^{26}$Al | 346° | 2° | 262 pc | Kretschmer et al (2013 – fig.9a) |
| | Masers | 346.5 | 2.5° | 327 pc | mean in Table 7 |
| | NIR star counts | 348° | 4° | 520 pc | mean in Table 10 |
| | Dust 870 μm | 343° | -1° | -131 pc | mean in table 9 |
| Start of Norma | $^{12}$CO at 8' | 020° | 0° | 0 pc, at 6.5 kpc | Vallée (2016 –Fig.2) |
| | NIR star counts | 019° | 1° | 113 pc | mean in Table 10 |
| | Sync. 6 GHz | 016° | 4° | 454 pc | Hayakawa et al (1981 – fig. 2b) |
| | Masers | 016.0 | 4° | 454 pc | mean in Table 7 |
| Scutum | $^{12}$CO at 8' | 032.9° | 0° | 0 pc, at 5 kpc | mean in Table 5 |
| | $^{26}$Al | 032° | 0.9° | 78 pc | Chen et al (1996 – fig.1) |
| | Thermal electron | 032° | 0.9° | 78 pc | Taylor & Cordes (1993 – fig.4) |
| | Sync. 408 MHz | 032° | 0.9° | 78 pc | Beuermann et al (1985 – fig.1) |
| | Old stars (1-8 μm) | 031.3° | 1.6° | 140 pc | mean in Table 10 |
| | $^{13}$CO | 031.3° | 1.6° | 140 pc | mean in Table 4 |
| | HI atom | 031.0° | 1.9° | 166 pc | mean in Table 8 |
| | Dust 240μm | 031° | 1.9° | 166 pc | Drimmell (2000 – fig. 1) |
| | Dust 870μm | 030.9° | 2.0° | 174 pc | mean in Table 9 |
| | 1.4GHz RRL | 030.8° | 2.1° | 183 pc | Hou & Han (2015 – table 1) |
| | Warm $^{12}$CO cores | 030° | 2.9° | 253 pc | Solomon et al (1985 – fig. 1b) |
| | FIR [CII] & [NII] | 030° | 2.9° | 253 pc | Steiman-Cameron et al (2010 – sect 2.1) |
| | HII complex | 029.6° | 3.3° | 288 pc | mean in Table 6 |
| | Dust 2.4μm | 029° | 3.9° | 340 pc | Hayakawa et al (1981 – fig. 2a) |
| | Masers | 028.5° | 4.4° | 384 pc | mean in Table 7 |
| | Dust 60μm | 026° | 6.9° | 602 pc | Bloemen et al (1990 – fig.5) |
| Sagit-tarius | Old stars (1-8 μm) | 055.0° | - 4.5° | -314 pc | mean in Table 10 |
| | HII complex | 050.6° | - 0.1° | -7 pc | mean in Table 6 |
| | HI atom | 050.6° | - 0.1° | -7 pc | mean in Table 8 |
| | $^{12}$CO at 8' | 050.5° | 0° | 0 pc, at 4 kpc | mean in Table 5 |
| | $^{13}$CO | 050.2° | 0.3° | 21 pc | mean in Table 4 |
| | Dust 240μm | 050° | 0.5° | 35 pc | Drimmell (2000 – fig. 1) |
| | Masers | 050.0° | 0.5° | 35 pc | mean in Table 7 |
| | FIR [CII] & [NII] | 050° | 0.5° | 35 pc | Steiman-Cameron et al (2010 – sect. 2.1) |
| | 1.4GHz RRL | 049.2° | 1.3° | 91 pc | Hou & Han (2015 – table 1) |
| | Dust 870μm | 049.1° | 1.4° | 98 pc | mean in Table 9 |
| | Warm $^{12}$CO cores | 049° | 1.5° | 105 pc | Solomon et al (1985 – fig. 1b) |
| | Thermal electron | 049° | 1.5° | 105 pc | Taylor & Cordes (1993 – fig.4) |
| | Sync. 408 MHz | 048° | 2.5° | 175 pc | Beuermann et al (1985 – fig. 1) |
| | $^{26}$Al | 046° | 4.5° | 316 pc | Chen et al (1996 – fig.1) |

Notes: (a): Published since 1980. Updating the earlier table 1 of Vallée (2014b).
(b): When there are 2 or more published reports for a given arm tracer in a given spiral arm, then a separate table is provided (Tables 3 to 10 here).
(c): Angular distance from the arm center, being positive towards the arm's inner edge (towards the Galactic Center), and negative in other direction (towards the galactic anti-center).
(d): Linear separation from the arm center ($^{12}$CO), after converting the angular separation at the arm distance from the sun; using 8.0 kpc for the distance of the Sun to the Galactic Center.

## Table 4 - Individually observed tangent longitude for the $^{13}CO$ J=1-0 tracer[(a)]

| Arm name | Tangent galactic longitude | Teles-cope HPBW | Survey name | Reference |
|---|---|---|---|---|
| Scutum | 030.5° | 46" | Galactic Ring | Hou & Han (2015- table 1) – reassessed [(b)] |
| | 032° | 3' | Bell Labs | Stark & Lee (2006 – fig.1, v= +95 km/s) |
| | 031.3° ±1.1° | mean and r.m.s. | | |
| Sagittarius | 049.4° | 46" | Galactic Ring | Hou & Han (2015- table 1) – reassessed [(b)] |
| | 051° | 3' | Bell Labs | Stark & Lee (2006 – fig.1, v= +60 km/s) |
| | 050.2° ±1.1° | mean and r.m.s. | | |

Notes:

(a): Published since 1980.

(b): Hou & Han (2015) reassessed the published $^{13}CO$ J=1-0 data from Jackson et al (2006).

## Table 5 - Indivdually observed tangent longitude for the $^{12}CO$ J=1-0 tracer[(a)]

| Arm name | Tangent Galactic longitude | Teles-cope HPBW | Survey name | Reference |
|---|---|---|---|---|
| Carina | 280° | 8.8' | Columbia | Alvarez et al (1990 – table 4) |
| | 280° | 8.8' | Columbia | Grabelsky et al (1987 – sect. 3.1.2) |
| | 281° | 8.8' | Columbia | Grabelsky et al (1988 – fig.4) |
| | 282° | 8.8' | Columbia | Bronfman et al (2000b – table 2) |
| | 282.0 | 8.8' | Columbia | Hou & Han (2015- table 1) – reassessed [(b)] |
| | 283° | 8.8' | Columbia | Bronfman et al (2000a – Section 3.4) |
| | 281.3° ±1.2° | mean and r.m.s. | | s.d.m of 0.5°, worth 44 pc at 5 kpc |
| Crux-Cen-Taurus | 308° | 8.8' | Columbia | Bronfman et al (2000a – sect. 3.4) |
| | 308° | 8.8' | Columbia | Bronfman (2008 – sect. 4) |
| | 309° | 8.4' | CfA | Dame & Thaddeus (2011 – fig.4) |
| | 309° | 8.8' | Columbia | Bronfman et al (2000b – table 2) |
| | 310° | 8.8' | Columbia | Alvarez et al (1990 – table 4) |
| | 310° | 8.8' | Columbia | Bronfman et al (1988 – fig. 7) |
| | 310° | 8.8' | Columbia | Bronfman et al (1989 – sect. 4) |
| | 310° | 8.8' | Columbia | García et al (2014 – table 3) |
| | 310° | 8.8' | Columbia | Grabelsky et al (1987 – sect. 3.1.2) |
| | 311.0° | 8.8' | Columbia | Hou & Han (2015- table 1) – reassessed [(b)] |
| | 309.5° ±1.0° | mean and r.m.s. | | s.d.m of 0.3°, worth 32 pc at 6 kpc |
| Norma | 328° | 8.8' | Columbia | Alvarez et al (1990 – table 4) |
| | 328° | 8.8' | Columbia | Bronfman et al (1988 – fig. 7) |
| | 328° | 8.8' | Columbia | Bronfman et al (1989 – sect. 4) |
| | 328° | 8.8' | Columbia | Bronfman (1992 – fig.6) |
| | 328° | 8.8' | Columbia | Bronfman et al (2000a – sect. 3.4) |
| | 328° | 8.8' | Columbia | Bronfman et al (2000b – table 2) |
| | 328° | 8.8' | Columbia | Bronfman (2008 - sect. 4) |
| | 328.3 | 8.8' | Columbia | Hou & Han (2015- table 1) – reassessed [(b)] |
| | 330° | 8.8' | Columbia | García et al (2014 – table 3) |
| | 330° | 8.8' | Columbia | Grabelsky et al (1987 – sect. 3.1.2) |
| | 328.4° ±0.8° | mean and r.m.s. | | s.d.m of 0.3°, worth 37 pc at 7 kpc |
| Start of Perseus | 336° | 8.8' | Columbia | Bronfman et al (1989 – sect. 4) |
| | 336° | 8.8' | Columbia | Bronfman (2008 – sect. 4) |
| | 336.7 | 8.8' | Columbia | Hou & Han (2015- table 1) – reassessed [(b)] |
| | 337° | 8.8' | Columbia | Alvarez et al (1990 – table 4) |
| | 337° | 8.8′ | Columbia | Bronfman et al (2000a – sect. 3.4) |
| | 337° | 8.8′ | Columbia | Bronfman et al (2000b - table 2) |
| | 337° | 8.8′ | Columbia | Dame & Thaddeus (2008 – sect. 1) |
| | 338° | 8.8′ | Columbia | García et al (2014 – table 3) |
| | 336.8° ±0.6° | mean and r.m.s. | | s.d.m of 0.3°, worth 42 pc at 8 kpc |
| Start of Sagittarius | 344° | 8.8′ | Columbia | Grabelsky et al (1987 – fig.6) |
| | 344° | mean | | |
| Scutum | 030.5° | 8.4′ | CfA | Hou & Han (2015- table 1) – reassessed (b) |
| | 031° | 8.4′ | CfA | Dame & Thaddeus (2011 – fig.4) |
| | 033° | 1.1′ | NRAO | Sanders et al (1985 – fig.5b) |
| | 034° | 7.5′ | Columbia | Dame et al (1986 – fig.9) |
| | 034° | 7.5′ | Columbia | Cohen et al (1980 – fig.3) |
| | 035° | 1.0′ | Texas | Chiar et al (1994 – Sect. 3) |
| | 032.9° ±1.8° | mean and r.m.s. | | s.d.m of 0.8°, worth 70 pc at 5 kpc |
| Sagittarius | 049.4 | 8.4′ | CfA | Hou & Han (2015- table 1) – reassessed (b) |
| | 051° | 7.5′ | Columbia | Cohen et al (1980 – fig.3) |
| | 051° | 7.5′ | Columbia | Dame et al (1986 – fig.9) |
| | 050.5° ±0.9° | mean and r.m.s. | | s.d.m of 0.5°, worth 36 pc at 4 kpc |

Notes:

(a): Published since 1980.

(b): Hou & Han (2015) reassessed the published $^{12}CO$ 1-0 data from Dame et al (2001).

## Table 6 – Individually observed tangent longitude for the HII region tracer[a]

| Arm name | Tangent galactic longitude | Range | Reference |
|---|---|---|---|
| Carina | 283.3° | radio-IR-opt. | Hou & Han (2015 – table 1) – reassessed [a] |
| | 284° | optical | Russeil (2003 – table 6) |
| | 284° | radio | Downes et al (1980 – fig.4) |
| | 283.8° ±0.3° | mean and r.m.s. | |
| Crux-Centaurus | 309° | optical | Russeil (2003 – table 6) |
| | 309° | radio | Downes et al (1980 – fig.4) |
| | 311.7° | radio-IR-opt. | Hou & Han (2015 – table 1) – reassessed [a |
| | 309.9° ±1.6° | mean and r.m.s. | |
| Norma | 323° | optical | Russeil (2003 – table 6) |
| | 328° | radio | Downes et al (1980 – fig.4) |
| | 328.1° | radio-IR-opt. | Hou & Han (2015 – table 1) – reassessed [a] |
| | 326.4° ±2.9° | mean and r.m.s. | |
| Start of Perseus | 337.2° | radio-IR-opt. | Hou & Han (2015 – table 1) – reassessed [a] |
| | 337.2° | mean | |
| Scutum | 025° | radio | Sanders et al (1985 – fig. 5a) |
| | 030.6° | radio-IR-opt. | Hou & Han (2015 – table 1) – reassessed [a] |
| | 031° | radio | Downes et al (1980 – fig.4) |
| | 032° | optical | Russeil (2003 – table 6) |
| | 029.6° ±3.1° | mean and r.m.s. | |
| Sagittarius | 046° | radio | Downes et al (1980 – fig.4) |
| | 049.4° | radio-IR-opt. | Hou & Han (2015 – table 1) – reassessed [a] |
| | 051° | optical | Russeil et al (2007 – fig.4) |
| | 056° | optical | Russeil (2003 – table 6) |
| | 050.6° ±4.2° | mean and r.m.s. | |

Notes:

(a): Published since 1980.

(b): Hou & Han (2015) reassessed the published HII regions data from Anderson et al (2014).

## Table 7 –Individually observed tangent longitude for the maser tracer[a]

| Arm name | Tangent Galactic longitude | Maser name | Reference |
|---|---|---|---|
| Carina | 284.5° | methanol | Hou & Han (2015 – table 1) – reassessed [b] |
| | 284.5° | mean | |
| Crux-Centaurus | 312.2° | methanol | Hou & Han (2015 – table 1) – reassessed [b] |
| | 312.2° | mean | |
| Norma | 329.3° | methanol | Hou & Han (2015 – table 1) – reassessed [b] |
| | 331.5° | methanol | Caswell et al (2011 – Sect. 4.6.2) |
| | 330.4° ±1.5° | mean and r.m.s. | |
| Start of Perseus | 337.0° | methanol | Hou & Han (2015 – table 1) – reassessed [b] |
| | 338° | methanol | Green et al (2011 – sect. 3.3.1) |
| | 338° | methanol | Green et al (2012 – Sect.2) |
| | 338° | methanol | Caswell et al (2011 – sect. 4.6.1) |
| | 337.8° ±0.5° | mean and r.m.s. | |
| Start of Sagittarius | 344° | $H_2O$ & methanol | Sanna et al (2014 – fig. 6) |
| | 349° | methanol | Green et al (2011 – table 1) |
| | 346.5° ±3.5° | mean and r.m.s. | |
| Start of Norma | 020° | methanol | Green et al (2011 – table 1) |
| | 012° | methanol | Green et al (2011 – table 1) |
| | 016.0° ±5.6° | mean and r.m.s. | |
| Scutum | 025° | $H_2O$ & methanol | Sanna et al (2014 – fig.6) |
| | 026° | methanol | Green et al (2012 – Sect.2) |
| | 028° | methanol | Green et al (2011 – sect. 3.3.1) |
| | 030° | meth., water | Reid et al (2014 – fig.1) |
| | 030.8° | methanol | Hou & Han (2015 – table 1) – reassessed [b] |
| | 031° | meth., water | Sato et al (2014 – fig3) |
| | 028.5° ±2.6° | mean and r.m.s. | |
| Sagittarius | 049.3° | methanol | Hou & Han (2015 – table 1) – reassessed [b] |

| | | |
|---|---|---|
| 049.6$^{o}$ | methanol | Pandian & Goldsmith (2007 – sect.4) |
| 050$^{o}$ | meth., water | Reid et al (2014 – fig.1) |
| 051$^{o}$ | meth., water | Wu et al (2014 – sect. 4.2) |
| 050.0$^{o}$ ±0.7$^{o}$ | mean and r.m.s. | |

Notes:

(a): Published since 1980

(b): Hou & Han (2015) reassessed the published methanol masers catalog from Hou & Han (2014).

## Table 8 - Individually observed tangent longitude for the HI atom tracer[(a)]

| Arm name | Tangent Galactic Longitude | Teles-cope HPBW | Survey name | Reference |
|---|---|---|---|---|
| Carina | 281.2° | 36' | Leiden | Nakanishi & Sofue (2016- fig.7)[(b)] |
| | 283.0° | 36' | LAB | Hou & Han (2015 – Table 1)[(c)] |
| | 282.1° ±1.3° | mean and r.m.s. | | |
| Crux-Centaurus | 309.3° | 36' | Leiden | Nakanishi & Sofue (2016- fig.7)[(b)] |
| | 310° | 15' – 36' | Hat Creek;Parkes | Englmaier & Gerhard (1999- table 1) |
| | 310.4° | 36' | LAB | Hou & Han (2015 – Table 1)[(c)] |
| | 309.9° ±0.6° | mean and r.m.s. | | |
| Norma | 328.0° | 36' | LAB | Hou & Han (2015 – Table 1)[(c)] |
| | 328.0° | 15' – 36' | Hat Creek;Parkes | Englmaier & Gerhard (1999- table 1) |
| | 328.4° | 36' | Leiden | Nakanishi & Sofue (2016- fig.7)[(b)] |
| | 328.1° ±0.2° | mean and r.m.s. | | |
| Start of Perseus | 336.8° | 36' | LAB | Hou & Han (2015 – Table 1)[(c)] |
| | 336.9° | 36' | Leiden | Nakanishi & Sofue (2016- fig.7)[(b)] |
| | 336.8° ±0.1° | mean and r.m.s. | | |
| Scutum | 029° | 15' – 36' | Hat Creek;Parkes | Englmaier & Gerhard (1999 – table 1) |
| | 030.8° | 36' | LAB | Hou & Han (2015 – Table 1)[(c)] |
| | 033.2° | 36' | Leiden | Nakanishi & Sofue (2016- fig.7)[(b)] |
| | 031.0° ±2.1° | mean and r.m.s. | | |
| Sagittarius | 050° | 15' – 36' | Hat Creek;Parkes | Englmaier & Gerhard (1999 – table 1) |
| | 050.8° | 36' | LAB | Hou & Han (2015 – Table 1)[(c)] |
| | 051.0° | 36' | Leiden | Nakanishi & Sofue (2016- fig.7)[(b)] |
| | 050.6° ±0.5° | mean and r.m.s. | | |

Notes:
(a): Published since 1980.
(b): Nakanishi & Sofue (2016) reassessed the published HI catalogs from Hartmann & Burton (1997) and Bajaja et al (2005); they added the CO survey of Dame et al (2001).
(c): Hou & Han (2015) reassessed the published HI catalog from Kalberla et al (2005).

## Table 9 - Individually observed tangent longitude for the dust 870μm trace[a]

| Arm name | Tangent galactic longitude | Teles-cope HPBW | Survey name | Reference |
|---|---|---|---|---|
| Carina | 284.2.0° | 19" | Atlasgal | Hou & Han (2015 – Table 1)[b] |
| | 284.2° | mean | | |
| Crux-Centaurus | 311° | 19" | Atlasgal | Beuther et al (2012 – fig.2) |
| | 311.7° | 19" | Atlasgal | Hou & Han (2015 – Table 1)[b] |
| | 311.4° ±0.5° | mean and r.m.s. | | |
| Norma | 327.2° | 19" | Atlasgal | Hou & Han (2015 – Table 1)[b] |
| | 332° | 19 " | Atlasgal | Beuther et al (2012 – fig.3) |
| | 329.6° ±3.4° | mean and r.m.s. | | |
| Start of Sagittarius | 343° | 19" | Atlasgal | Beuther et al (2012 – fig.3) |
| | 343° | mean | | |
| Start of Perseus | 337.5° | 19" | Atlasgal | Hou & Han (2015 – Table 1)[b] |
| | 338° | 19" | Atlasgal | Beuther et al (2012 – fig.3) |
| | 337.8° ±0.3° | mean and r.m.s. | | |
| Scutum | 031° | 19" | Atlasgal | Beuther et al (2012 – fig.3) |
| | 030.7° | 19" | Atlasgal | Hou & Han (2015 – Table 1)[b] |
| | 030.9° ±0.2° | mean and r.m.s. | | |
| Sagittarius | 049.2° | 19" | Atlasgal | Hou & Han (2015 – Table 1)[b] |
| | 049° | 19" | Atlasgal | Beuther et al (2012 – fig.3) |
| | 049.1° ±0.2° | mean and r.m.s. | | |

Notes:
(a): Published since 1980.
(b): Hou & Han (2015) reassessed the published HI catalogs from Contreras et al (2013) and Csengeri et al (2014).

## Table 10 - Individually observed tamgent longitude for the 1µm to 8µm old stars tracer [a]

| Arm name | Tangent galactic longitude | Teles-cope HPBW | Survey name | Reference |
|---|---|---|---|---|
| Crux-Centaurus | 307° | 21’ | COBE K-band 2µm | Drimmel (2010 – Sect.3) |
| | 307.5° | 1.2;2” | GLIMPSE; 2MASS | Hou & Han (2015 – Table 1)[b] |
| | 307.3° ±0.4° | mean and r.m.s. | | |
| Start of Perseus | 338.3° | 1.2”;2” | GLIMPSE; 2MASS | Hou & Han (2015 – Table 1)[b] |
| | 338.3° | mean | | |
| Start of Sagittarius | 348° | 21’ | J,H,K COBE | Benjamin (2008 – Fig.2) |
| | 348° | mean | | |
| Start of Norma | 019° | 21’ | J,H,K COBE | Benjamin (2008 – Fig.2) |
| | 019° | 1.2;2” | 4.5µm GLIMPSE | Benjamin (2008 – Fig.2) |
| | 019° | mean | | |
| Scutum | 030° | 21’ | COBE K-band 2µm | Drimmel (2010 – Sect.3) |
| | 032.6° | 1.2”;2” | GLIMPSE; 2MASS | Hou & Han (2015 – Table 1)[b] |
| | 031.3° ±1.8° | mean and r.m.s. | | |
| Sagittarius | 055.0° | 1.2”;2” | GLIMPSE; 2MASS | Hou & Han (2015 – Table 1)[b] |
| | 055.0° | mean | | |

Notes:
(a): published since 1980.
(b): Hou & Han (2015) reassessed the published old star catalogs from Benjamin et al (2003; GLIMPSE) and Skrutskie et al (2006; 2MASS).